\input psfig.sty
\documentstyle[twocolumn]{mn}
\newcommand{\mincir}{\raise -2.truept\hbox{\rlap{\hbox{$\sim$}}\raise5.truept
\hbox{$<$}\ }}
\newcommand{\magcir}{\raise -2.truept\hbox{\rlap{\hbox{$\sim$}}\raise5.truept
\hbox{$>$}\ }}
\newcommand{\minmag}{\raise-2.truept\hbox{\rlap{\hbox{$<$}}\raise 6.truept\hbox
{$>$}\ }}
\newcommand{\be}{\begin{equation}}
\newcommand{\ee}{\end{equation}}
\newcommand{\ba}{\begin{eqnarray}}
\newcommand{\ea}{\end{eqnarray}}
\newcommand{\brr}{\begin{array}}
\newcommand{\err}{\end{array}}
\newcommand{\bc}{\begin{center}}
\newcommand{\ec}{\end{center}}

\newcommand{\hm}{\,h^{-1}\,{\rm Mpc}}

\title{The Epoch of Structure Formation in Blue Mixed Dark Matter Models}
\author[S. Borgani, F. Lucchin, S. Matarrese and L. Moscardini]
{Stefano Borgani$^{1,2}$, Francesco Lucchin$^{3}$,
Sabino Matarrese$^{4}$ and Lauro Moscardini$^{3}$ \\
$^1$ INFN -- Sezione di Perugia, c/o Dipartimento di Fisica dell'Universit\`a,
via A. Pascoli, I--06100 Perugia, Italy \\
$^2$SISSA -- International School for Advanced Studies,
via Beirut 2--4, I--34013 Trieste, Italy \\
$^3$Dipartimento di Astronomia, Universit\`a di Padova,
vicolo dell'Osservatorio 5, I--35122 Padova, Italy \\
$^4$Dipartimento di Fisica {\it Galileo Galilei}, Universit\`a di Padova,
via Marzolo 8, I--35131 Padova, Italy }

\begin{document}

\maketitle

\begin{abstract}
Recent data on the high--redshift abundance of damped Ly$\alpha$ systems 
are compared with theoretical predictions for `blue' (i.e. $n>1$) Mixed
Dark Matter models. The results show that decreasing the hot component fraction
$\Omega_\nu$ and/or increasing the primordial spectral index $n$ implies an
earlier epoch of cosmic structure formation. However, we also show that 
varying $\Omega_\nu$ and $n$ in these directions makes the models barely 
consistent with the observed abundance of galaxy clusters. Therefore, 
requiring at the same time observational constraints on damped Ly$\alpha$ 
systems and cluster abundance to be satisfied represents a 
challenge for the Mixed Dark Matter class of models.

\end{abstract}

\section{Introduction}
Since a long time observations of high--redshift objects have become a
potentially powerful constraint for models of cosmic structure formation. The
availability of statistically reliable samples of quasars allowed to address
this problem in a quantitative way in the framework of the Cold Dark Matter
cosmogony (Efstathiou \& Rees 1988; Haehnelt 1993). Moreover, the comparison of
predictions and observations of quasar abundance at different redshifts 
has been used as a test for model reliability (e.g. Nusser \& Silk 1993; 
Pogosyan \& Starobinsky 1993). 

Recently, damped Ly$\alpha$ systems (DLAS) have been recognized as a promising
way to 
trace the presence of high redshift collapsed structures,
thanks to the possibility of identifying them as protogalaxies and to their
detectability at 
high $z$ (see Wolfe 1993 for a comprehensive review). DLAS are seen as wide
absorbtion features in quasar spectra. The associated absorbing systems have a
neutral hydrogen column density $\ge 10^{20}\,$ cm$^{-2}$. The rather large
abundance of DLAS makes it possible to compile reliable statistical samples
(Lanzetta 1993; Lanzetta, Wolfe \& Turnshek 
1995; Storrie--Lombardi et al. 1995;
Wolfe et al. 1995). Once the
parameters of the Friedmann background are specified, observations of DLAS can
be turned into the value of the cosmological density parameter $\Omega_{g}$
contributed by the neutral gas, which is associated with DLAS. It turns out that
at $z\sim 3$ this quantity is comparable to the mass density of visible matter
in nearby galaxies, thus suggesting that DLAS trace a population of galaxy
progenitors. 

Based on the APM QSO 
catalogue, Storrie--Lombardi et al. (1995) recently presented
the most extended DLAS sample up-to-date, covering the range $2.8 < z < 4.4$.
In the following of this paper we will consider their highest redshift data as
the most constraining ones and we will compare them with model predictions. 

Several authors (Subramanian \& Padmanabhan 1994; Mo \&
Miralda--Escud\'e 1994; Kauffmann \& Charlot 1994; Ma \& Bertschinger 1994)
have recently claimed that the large value of $\Omega_g$
observed at $z~\magcir 3$ is incompatible with predictions of the
Mixed (i.e.
cold+hot) Dark Matter (MDM) model with spectral index $n=1$ and
$\Omega_\nu\simeq 0.3$ contributed by
one species of massive neutrinos 
and $\Omega_b=0.1$ for the baryon fractional density 
(Klypin et al. 1993; Nolthenius, Klypin \& Primack 1995). 
Klypin et al. (1995) reached substantially the same
conclusions about this model, but emphasized two relevant points: ({\it i}) any
theoretical prediction is very sensitive to the choice of the parameters of the
model needed to obtain $\Omega_g$ from a given power--spectrum; ({\it ii}) 
slightly
lowering $\Omega_\nu$ to 20--25\% keeps MDM into better agreement with DLAS
data, independently of whether the hot component is given by one or two massive
neutrino species (see also Primack et al. 1995).

\begin{figure*}
\centerline{
\psfig{figure=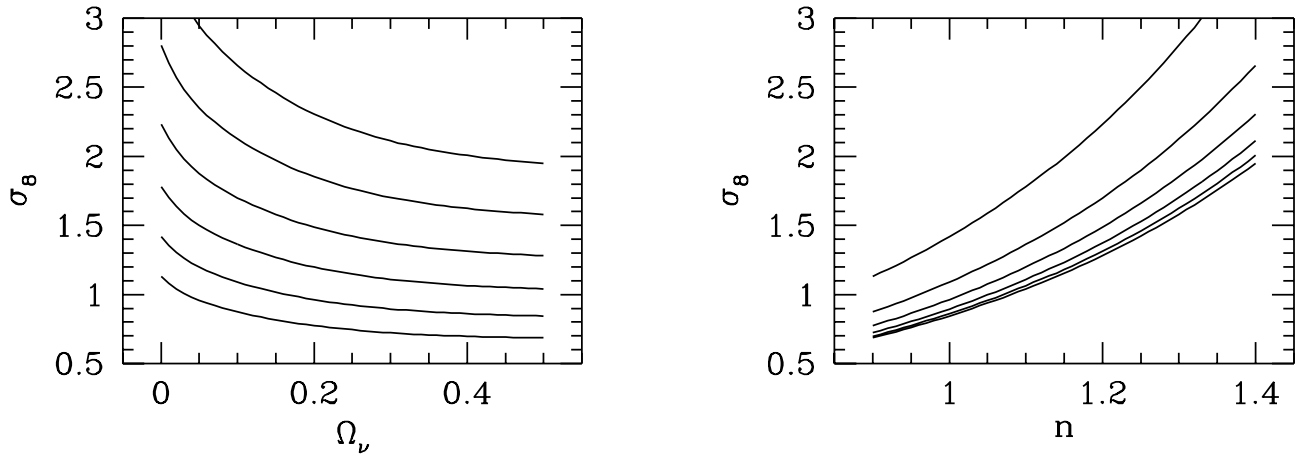,height=7cm}
}
\caption{The r.m.s. fluctuation amplitude within a top--hat sphere of 
$8\hm$, $\sigma_8$, for {\em COBE} normalization (see text).
Left panel: dependence on the hot component fraction
$\Omega_\nu$; different lines refer to $n=0.9,1,1.1,1.2,1.3,1.4$ from bottom to
top. Right panel: dependence on the spectral index $n$; different lines refer
to $\Omega_\nu=0,0.1,0.2,0.3,0.4,0.5$ from top to bottom.} 
\label{fi:sig8}
\end{figure*}

A possible alternative, still in the framework of MDM, is to consider `blue'
($n>1$) primordial spectra of density fluctuations. The advantage of these
models is an anticipation of the epoch of structure formation due to the higher
small--scale power. The choice of blue spectra was originally suggested by 
the analysis of Cosmic Microwave Background anisotropies on scales larger than 
$1^\circ$ (e.g. Devlin et al. 
1994; Hancock et al. 1994; Bennett et al. 1994). Possible indications for blue 
spectra comes from large bulk flows (Lauer \& Postman 1994; see however 
Riess, Press \& Kirshner 1995; Branchini \& Plionis 1995) and large 
voids in the
galaxy distribution (Piran et al. 1993). In recent years many authors have
pointed out that the inflationary dynamics can easily account for the origin
of blue perturbation spectra (Liddle \& Lyth 1993; Linde 1994; Mollerach, 
Matarrese \& Lucchin 1994; 
Copeland et al. 1994), in particular in the framework of the so--called hybrid
models. Recently Lucchin et al. (1995), using linear theory and N--body
simulations, performed an extended analysis of the large--scale structure
arising from blue MDM (BMDM) models: the most interesting advantage of these
models is the increase of the galaxy formation redshift: for instance, taking  
$\Omega_\nu=0.3$ one has for the redshift of non--linearity on galactic scale
($M=10^{12}~{\rm M}_\odot$) 
$z_{nl} \approx 1.9$ if $n=1.2$ and $z_{nl}\approx 0.6$ if
$n=1$. The same class of models has been tested against observational data,
using linear theory predictions, by Dvali, Shafi \& Schaefer (1994) and
Pogosyan \& Starobinsky (1995a). 

\begin{figure*}
\centerline{
\psfig{figure=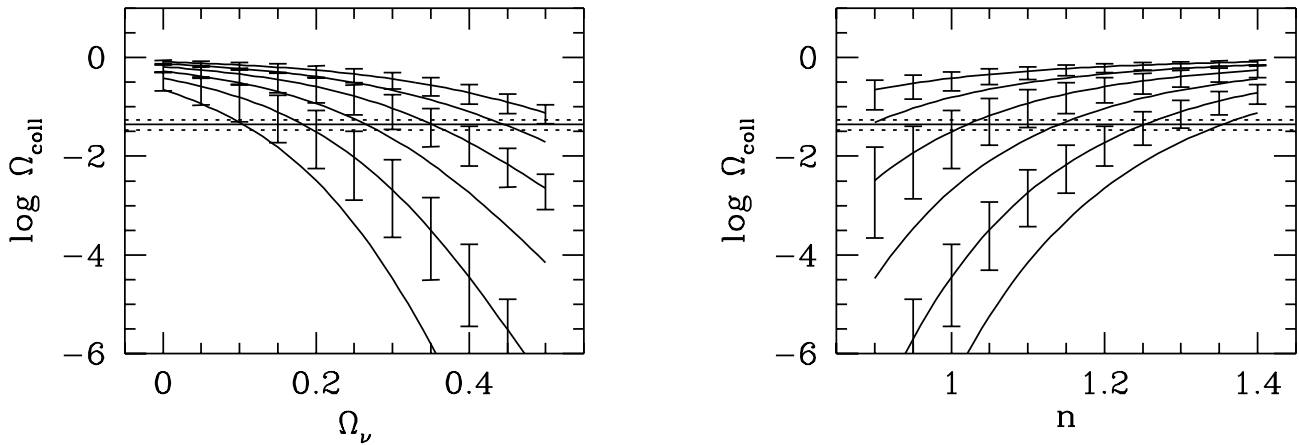,height=7cm}
}
\caption{The fractional matter density within collapsed structures
$\Omega_{coll}$ at redshift $z=4.25$, when $\delta_c=1.5$ and $M=10^{11}~{\rm
M}_\odot$ are assumed. Left panel: dependence on the hot component fraction
$\Omega_\nu$; different lines refer to $n=0.9,1,1.1,1.2,1.3,1.4$ from bottom to
top. Right panel: dependence on the spectral index $n$; different lines refer
to $\Omega_\nu=0,0.1,0.2,0.3,0.4,0.5$ from top to bottom. Error bars show the
effect of taking $M=10^{10}~{\rm M}_\odot$ and $10^{12}~{\rm M}_\odot$. The
horizontal lines refer to the observational data by Storrie--Lombardi et al.
(1995) with the corresponding uncertainties. 
} 
\label{fi:omco}
\end{figure*}

In this
work we will compare BMDM model predictions, for different values of
$\Omega_\nu$ and $n$ (for the sake of comparison we also consider the 
$0.9\le n<1$ tilted models), with the observed DLAS abundance. Furthermore,
we will also discuss the implications of the observed abundance of galaxy 
clusters for BMDM models. In fact, increasing $n$ for a fixed $\Omega_\nu$
and fixed normalization to {\it COBE} rises up the value of $\sigma_8$, the 
r.m.s. fluctuation within a top--hat
sphere of  $8\hm$ radius, which is constrained  by the observed abundance
of galaxy clusters to be $\sigma_8\simeq 0.6$ (White, Efstathiou \& Frenk
1993). 

\section{The method} 
In order to connect our model predictions to DLAS observables, let us define
$\Omega_{coll}(z)$ as the fractional matter density within collapsed structures
at redshift $z$. Therefore
\be
\Omega_{coll}(z)\,=\,{\Omega_g (z)\over \Omega_b f_g}\ ,
\label{eq:ocoll}
\ee
where $\Omega_{b}$ is the fractional baryon density (since $h=0.5$ is
assumed throughout the paper, we take $\Omega_b=0.05$ according to standard
primordial nucleosynthesis; see, e.g., Reeves 1994)  and $f_g$ is the
fraction of the HI gas, which is involved in DLAS. Although the observed
decrease of $\Omega_g$ with redshift for $z~\mincir 3.5$  is usually
considered as an indication of gas consumption into stars (e.g. Lanzetta et
al. 1995; Wolfe et al. 1995), the actual value of $f_g$ at the high
redshift we are interested in is not clear. In any case, since $\Omega_g$
at such high redshifts is quite similar to the fractional density
contributed by visible matter in present--day normal galaxies, we expect
$f_g$ not to be a particularly small number. In the following we will show
results based both on $f_g=0.5$ and 1. 

Taking $h=0.5$ and an Einstein--de Sitter universe, the data at $z=4.25$ from
Storrie--Lombardi et al. (1995) turn into $\Omega_{coll}=(8.8\pm 2.0)\times
10^{-2}$ and $(4.4\pm 1.0)\times 10^{-2}$ for $f_{g}=0.5$ and 1, respectively. 

From the theoretical side, the Press \& Schechter (1974) approach gives a
recipe to compute the contribution to the cosmic density due to the matter
within collapsed structures of mass $M$ at redshift $z$: 
\be
\Omega_{coll}(M,z)\,=\,{\rm erfc}\left({\delta_c\over \sqrt2\sigma_M(z)}\right)
\,. \label{eq:prsc}
\ee
The above expression assumes Gaussian fluctuations and $\delta_c$ is the
linearly extrapolated density contrast for the collapse of a perturbation;
$\sigma_M$ is the r.m.s. fluctuation at the mass--scale $M$, where 
\be
M\,=\,(2\pi R^2)^{3/2} \rho
\label{eq:mr}
\ee
for the Gaussian window that we will assume in the following. Here, $\rho$ 
is the average matter density, which is taken to have the critical value.

As for the mass of the structures hosting DLAS, it has been argued that, since
the high column density of the absorber is typical of large disks of luminous
galaxies, DLAS should be located within massive structures of $\sim
10^{12}~{\rm M}_\odot$. However, it is not clear at all whether the properties
of present--day galaxies can be extrapolated to their high--redshift
progenitors. Therefore, we prefer to leave open the possibility that DLAS are
hosted within smaller structures. It is clear that, when a model is in trouble 
in accounting for the DLAS abundance if the hosting structure is a dwarf galaxy
($M\sim 10^{10}~{\rm M}_\odot$), it would certainly be ruled out if more
massive protogalaxies are required. 

Linear theory for the top--hat spherical collapse predicts $\delta_c=1.69$.
However, effects of non--linearity as well as asphericity of the collapse
could cause significant deviations from
this value. Klypin et al. (1995) estimated the halo abundance at different
redshifts from high mass resolution N--body simulations. By using the 
Gaussian window, they found a good
agreement with the Press--Schechter expression for values as low as
$\delta_c=1.3$--1.4 (see also Efstathiou \& Rees 1988). On the other hand, Ma \&
Bertschinger (1994) found that for $\Omega_\nu=0.3$ and a top--hat window 
$\delta_c\simeq 1.8$ is always required, which corresponds to $\delta_c
\simeq 1.7$ for the Gaussian window. In the
following we prefer to show results based on $\delta_c=1.5$ but, due to the
previous uncertainties, we will discuss also the effect of different choices for
$\delta_c$.

\begin{figure*}
\centerline{
\psfig{figure=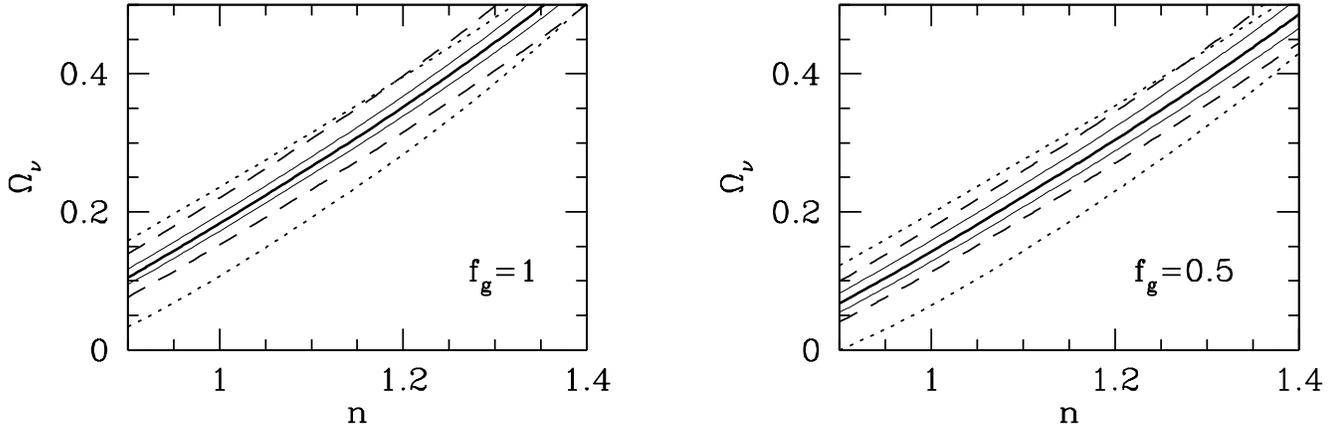,height=7cm}
}
\caption
{The models in the $\Omega_\nu$ -- $n$ plane reproducing the observed
$\Omega_{coll}$, taking $f_g=1$ (left panel) and $f_g=0.5$ (right panel). The
heavy solid curve is for $\delta_c=1.5$ and $M=10^{11}~{\rm M}_\odot$; lighter
solid curves refer to the observational uncertainties. Dashed and dotted lines
show the effect of varying $\delta_c$ and $M$, respectively (see the text for
the assumed values).} 
\label{fi:ondl}
\end{figure*}

For the MDM transfer function we take the fit obtained by Pogosyan \&
Starobinsky (1995a), which provides a continuous dependence on the 
fractional density $\Omega_\nu$
contributed by one massive neutrino.
As for the {\em cold} part of the transfer function, we use the Cold Dark Matter
expression by Efstathiou, Bond \& White (1992), with the shape parameter
$\Gamma=\Omega_\circ h\exp(-2\Omega_b)$, according to the prescription of
Peacock \& Dodds (1994), to account for the baryonic component. We varied
$\Omega_\nu$ in the interval $0\le \Omega_\nu \le 0.5$. We assume for the
primordial (post inflationary) power--spectrum $P(k)\propto k^n$ with $0.9\le
n\le 1.4$. 
Each model is normalized to the 9--th multipole component of the {\it
COBE} DMR two--year data, $a_9=8.2$, which has been shown to be independent 
of the $n$ value to a good accuracy (G\'orski et al. 1994). 

In Figure \ref{fi:sig8} we plot the resulting $\sigma_8$ value. As for the
$\Omega_\nu$ dependence, results are plotted in the left panel for $n=0.9 -
1.4$ going from lower to upper curves with steps of $0.1$. In a similar
fashion, in the right panel we plot the $n$ dependence. Going from higher
to lower curves, we plot results for $\Omega_\nu=0 - 0.5$ with steps of
0.1. As expected, $\sigma_8$ is an increasing function of $n$, while it 
decreases with $\Omega_\nu$.

\section{Discussion}
The results of our analysis on DLAS are summarized in Figure \ref{fi:omco},
where we plot $\Omega_{coll}$, estimated at $z=4.25$, as a function of
$\Omega_\nu$ (left panel) and of $n$ (right panel), after assuming
$\delta_c=1.5$ and $M=10^{11}~{\rm M}_\odot$. In each panel, different
curves are for the same choice of parameters as in Figure \ref{fi:sig8}.
Upper and lower error bars show the effect of taking $M=10^{10}~{\rm
M}_\odot$ and $10^{12}~{\rm M}_\odot$, respectively. The horizontal solid
line is the observational result with the corresponding uncertainties
(dotted lines), which is obtained by converting the $\Omega_g$ value, as
reported by Storrie--Lombardi et al. (1995) at $z=4.25$, to $\Omega_{coll}$
according to eq.(\ref{eq:ocoll}) with $f_g=1$. 

Figure \ref{fi:ondl} 
shows in the $\Omega_\nu$ -- $n$ plane the models which reproduce the
observed $\Omega_{coll}$, taking $f_g=1$ (left panel) and $f_g=0. 5$ (right
panel). The heavy solid curve corresponds to $\delta_c=1.5$ and $M=10^{11}~{\rm
M}_\odot$, with the lighter curves delimiting the observational uncertainties.
Upper and lower dashed lines show the effect of varying $\delta_c$ to 1.3 and
1.7, respectively. Upper and lower dotted curves refer to $M=10^{10}$ and
$10^{12}~{\rm M}_\odot$, respectively. The overall result that we get is
that decreasing the hot component fraction $\Omega_\nu$ and/or increasing
the primordial spectral index $n$ implies an earlier formation of cosmic
structures. 

It should be noted that realistic observational uncertainties should be
larger than the error bars reported by Storrie--Lombardi et al. (1995),
since they do not include any systematic observational bias. Recently, 
Bartelman \& Loeb (1995) emphasized the role of the amplification bias, due 
to DLAS gravitational lensing of QSOs, in the DLAS detection. They pointed 
out that {\bf (a)} lensing effects bias upwards the $\Omega_g(z)$ value by 
an amount depending on the parameters of the Friedmann background, as 
well as on the redshift; {\bf (b)} the observed absorber sample may be 
biased toward larger values of their internal line--of--sight velocity 
dispersions leading to an overestimate of the total absorber mass. 
Therefore, both effects go in the direction of alleviating the galaxy formation
redshift problem.
On the other hand, Fall \& Pei (1995) detailed the consequences of dust 
absorbtion in DLAS. They argued that dust obscuration causes incompleteness 
in the optically selected quasar samples, and, therefore, in the DLAS 
samples as well. In this case, the resulting $\Omega_g(z)$ is biased 
downwards by an amount depending on the model for the DLAS chemical 
evolution.

It is however clear that, even taking
the observational results at face value with their small error bars, the rather
poor knowledge of the parameters entering in the Press--Schechter prediction
for $\Omega_g$ (i.e. $\delta_c$, $M$ and $f_g$) makes it difficult to put
stringent constraints on $\Omega_\nu$ and $n$. 

For instance, if one takes $1.3~\mincir \delta_c~\mincir 1.5$, as suggested by
several N--body simulations (e.g. Efstathiou \& Rees 1988; Klypin et al. 1995)
and analytical considerations on the Press--Schechter approach (e.g. Jain \&
Bertschinger 1994), $\Omega_\nu \simeq 0.2$ and $n=1$ would be allowed for
$M\sim 10^{10} - 10^{11}~{\rm M}_\odot$, unless $f_g$ is sensibly below unity.
On the other hand, blueing the spectrum to $n=1.2$ increases the allowed hot
fraction to $\Omega_\nu ~\magcir 0.4$, unless $\delta_c\simeq 1.7$ or $M\simeq
10^{12}~{\rm M}_\odot$ are taken. In order to more tightly constrain the
models, a better understanding of galaxy formation through hydrodynamical
simulations would be needed to clarify what DLAS actually are. This would
provide more reliable values for $\delta_c$, $M$ and $f_g$. 

It is however clear that those models which fit the data at $z\simeq 4$
need also to be tested against present--day observables. One of these tests
is represented by the abundance of galaxy clusters, which has been shown to
represent a powerful constraint for dark matter 
models (e.g., White et al. 1993). 
In the Press \& Schechter (1974) approach, the number of density of 
clusters with mass above $M$ is given by
\be
N(>M)\,=\,\int_M^\infty n(M')\,dM'\,,
\label{eq:abo}
\ee
where
\be
n(M)\,dM~=~{\delta_c\over \alpha \sqrt{2\pi}}
\int_R^\infty{\eta(R)\over \sigma(R)}\,\exp\left(-{\delta_c^2\over 2\sigma^2
(R)}\right)\,{dR\over R^2}
\label{eq:nm}
\ee
is the average cluster number density with mass in the range $[M,M+dM]$. In 
the above expression, the quantities
\ba
\sigma^2(R) & = & {1\over 2\pi^2}\int k^2\,P(k)\,W^2(kR)\,dk \ , \nonumber \\
\eta(R) & = & {1\over 2\pi^2 \sigma^2(R)}\int k^4\,P(k)\,
{dW^2(kR)\over d(kR)}\,{dk\over kR}\,
\label{eq:etsi}
\ea
convey the information about the power--spectrum. As before, we use a 
Gaussian window for $W(kR)$, so that $\alpha=(2\pi)^{3/2}$ in 
eq.(\ref{eq:nm}) and the mass $M$ 
is related to the scale $R$ according to eq.(\ref{eq:mr}). 
Klypin \& Rhee (1994) found that $\delta_c\simeq 1.5$ for 
their MDM cluster N--body simulations, with $\Omega_\nu=0.3$ and Gaussian 
filter (see Borgani et al. 1995, for the dependence of the cluster mass 
function on 
$\delta_c$ for different dark matter models).

As for observational data, White et al. (1993) estimated a cluster
abundance of about $5\times 10^{-7}~{\rm Mpc}^{-3}$ for masses exceeding
$M=8.4\times 10^{14}~{\rm M}_\odot$ using $X$--ray data. Biviano et al.
(1993) based their analysis on observed cluster velocity dispersion and
obtained an abundance of about $7.5\times 10^{-7}~{\rm Mpc}^{-3}$ for clusters
exceeding the above mass limit. 

\begin{figure*}
\centerline{
\psfig{figure=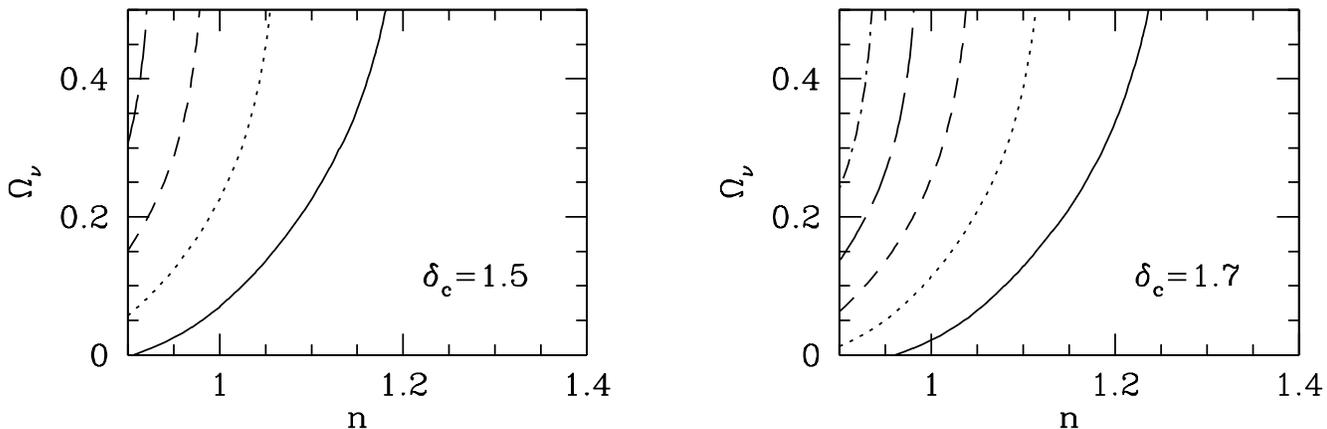,height=7cm}
}
\caption
{The models in the $\Omega_\nu$ -- $n$ plane corresponding to different 
values for the cluster abundance, $N(>M)$, taking $M=8.4\times 
10^{14}~{\rm M}_\odot$. 
Left and right panels correspond to assuming $\delta_c=1.5$ 
and $\delta_c=1.7$, respectively. Solid, dotted, 
short--dashed, long--dashed and dot--dashed curves (partly absent in the 
left panel) are for $\log N(>M)=-5.25,-5.5,-5.75,-6.,-6.25$,
respectively. 
For comparison, the observational results by White et al. (1993) and by Biviano 
et al. (1993) give $\log N(>M)\simeq -6.3$ and $-6.1$, respectively.
} 
\label{fi:clab}
\end{figure*}

In Figure \ref{fi:clab} we plot the $\Omega_\nu$ -- $n$ relations for 
different values of $N(>M)$ (see caption) 
and taking $M=8.4\times 10^{14}~{\rm M}_\odot$.
Quite remarkably, for $\delta_c=1.5$ (left panel) no values of $\Omega_\nu$ 
and $n$ in the considered ranges give a cluster abundance as low as the 
observational ones. For instance, assuming $(\Omega_\nu,n)=(0.2,1)$ it is 
$N(>M)\simeq 3.4\times 10^{-6}~{\rm Mpc}^{-3}$, while 
$N(>M)\simeq 7.1\times 10^{-6}~{\rm Mpc}^{-3}$ for $(\Omega_\nu,n)=(0.3,1.2)$.
In general, lowering $\Omega_\nu$ 
and/or increasing $n$, as suggested by the DLAS analysis, makes the 
disagreement even worse. This result agrees with the expectation that 
$\sigma_8\simeq 0.6$ (White et al. 1993) to reproduce the observed cluster 
abundance, while in general larger normalizations are required by our 
models (cf. Figure \ref{fi:sig8}). Increasing $\delta_c$ to 1.7 alleviates 
to some extent the disagreement. However, even in this case, in order to 
reproduce the observational $N(>M)$ one should have $\Omega_\nu\magcir 0.2$ 
and $n\mincir 0.95$, with the limiting values ($\Omega_\nu,n) =(0.2,0.9)$ 
being only marginally consistent with the DLAS constraints corresponding to 
most optimistic choice of the parameters (cf. the upper dotted curve in the 
left panel of Figure 2). This general problem of the models in reproducing 
the cluster abundance was also recognized by Pogosyan \& Starobinsky 
(1995a). The point is also discussed by Lucchin et al. (1995).

Although systematic observational uncertainties could well affect the 
determination of cluster masses from both $X$--ray and velocity dispersion 
data, it is not clear whether they can justify the order of magnitude (or 
even more) discrepancy between the data and those models which would have been 
preferred on the ground of DLAS constraints. 

A first possibility to alleviate this problem would be to increase the
baryon fraction $\Omega_b$. On one hand, this has the effect of
lowering the small--scale fluctuation amplitude and, therefore, $\sigma_8$.
On the other hand, this fluctuation suppression should not damage too
seriously DLAS predictions on $\Omega_g$, since this effect is partly
compensated by a larger denominator in eq.(\ref{eq:ocoll}).
For instance, taking $\delta_c=1.5$ for the $(\Omega_\nu,n)=(0.2,0.9)$ model
the cluster abundance changes from $n(>M)\simeq 1.4\times 10^{- 6}$ Mpc$^{-3}$
to $n(>M)\simeq 8.2\times 10^{- 7}$ Mpc$^{-3}$ when passing from 5\% to 10\% of
baryonic fraction. However, such an effect turns out not to be effective in
reconciling with observational data those models which largely overproduce
clusters. Indeed, even for $\delta_c=1.7$, $n(>M)$ for
$(\Omega_\nu,n)=(0.3,1.2)$ drops only from $7.1\times 10^{-6}$ Mpc$^{-3}$ to
$2.5\times 10^{-6}$ Mpc$^{-3}$ when passing from 5\% to 20\% of baryonic
fraction, the second value already being largely inconsistent with 
the primordial nucleosynthesis predictions. 

A further possibility is sharing the hot component between more than one
massive neutrino species (Primack et al. 1995; Pogosyan \& Starobinsky
1995b; Babu, Schaefer \& Shafi 1995). The subsequent variation of the
neutrino free--streaming has been shown to decrease $\sigma_8$ to an
adequate level, without significantly affecting results at the galactic
scale, which is relevant for DLAS. 

As a general conclusion we would stress the effectiveness of putting
together different kinds of observational constraints to restrict the range
of allowed models. As we have shown, the effect of blueing the primordial
perturbation spectrum goes in the  direction of increasing the redshift of
structure formation. However, this also increases the r.m.s. fluctuation on the
cluster mass scale to a dangerous level. Deciding whether the narrowing of
the allowed region of the $\Omega_\nu$ -- $n$ plane points towards the 
selection of the best model or towards ruling out the entire class of models 
requires a clarification of both the observational situation and of how models 
have to be compared to data. 

\section*{Acknowledgments} The Italian MURST is acknowledged for partial
financial support. SB wishes to thank Joel Primack for useful discussions.

\end{document}